\documentclass[journal]{IEEEtran}
\usepackage{soul} % in preamble
\usepackage{xcolor} % in preamble
\sethlcolor{yellow}
\usepackage{cite}
\usepackage{amsmath,amssymb,amsfonts}
\usepackage{algorithmic}
\usepackage{graphicx}
\usepackage{textcomp}
\usepackage{comment}
\usepackage[font=footnotesize,labelfont=bf,textfont=normalfont]{caption}
\usepackage{multirow}
\usepackage{float} % Ensures strict placement
\usepackage[colorlinks=true, linkcolor=blue, citecolor=blue, urlcolor=blue]{hyperref}
\usepackage{subfig} % For \subfloat command
\def\BibTeX{{\rm B\kern-.05em{\sc i\kern-.025em b}\kern-.08em
    T\kern-.1667em\lower.7ex\hbox{E}\kern-.125emX}}
\begin{document}
\title{AI-Driven Phase-Shifted Carrier Optimization for Cascaded Bridge Converters, Modular Multilevel Converters, and Reconfigurable Batteries}

\author{Amin Hashemi-Zadeh, \IEEEmembership{Student Member, IEEE,}
       Nima Tashakor, \IEEEmembership{Senior Member, IEEE,} Sandun Hettiarachchi, \IEEEmembership{Student Member, IEEE}, and
 Stefan Goetz, \IEEEmembership{Member, IEEE}% <-this % stops a %space
\thanks{Amin Hashemi-Zadeh and Nima Tashakor contributed equally to this work.}
%\thanks{Amin Hashemi-Zadeh and Sandun Hettiarachchi are with the Department of Electrical and Computer Engineering, Rheinland-Pfälzische Technische Universität (RPTU), 67663 Kaiserslautern, Germany (e-mail: aminhashemi.zadeh@rptu.de; don@rptu.de).}
%\thanks{Nima Tashakor and Stefan Goetz are with the Department of Electrical and Computer Engineering, Duke University, Durham, NC 27708 USA (e-mail: nima.tashakor@duke.edu; stefan.goetz@duke.edu).}
}
\maketitle

\begin{abstract}
Phase-shifted carrier pulse-width modulation (PSC-PWM) is a widely adopted scheduling algorithm in cascaded bridge converters, modular multilevel converters, and reconfigurable batteries. However, non-uniformed pulse widths for the modules with fixed phase shift angles lead to significant ripple current and output-voltage distortion. Voltage uniformity instead would require optimization of the phase shifts of the individual carriers. However, the computational burden for such optimization is beyond the capabilities of any simple embedded controller.
This paper proposes a neural network that emulates the behavior of an instantaneous optimizer with significantly reduced computational burden. The proposed method has the advantages of stable performance in predicting the optimum phase-shift angles under balanced battery modules with non-identical modulation indices without requiring extensive lookup tables, slow numerical optimization, or complex controller tuning. With only one (re)training session for any specified number of modules, the proposed method is readily adaptable to different system sizes. Furthermore, the proposed framework also includes a simple scaling strategy that allows a neural network trained for fewer modules to be reused for larger systems by grouping modules and adjusting their phase shifts. The scaling strategy eliminates the need for retraining.
Large-scale assessment, simulations, and experiments demonstrate that, on average, the proposed approach can reduce the current ripple and the weighted total harmonic distortion by up to 50\,\% in real time and is 100 to 500 thousand times faster than a conventional optimizer (e.g., genetic algorithms), making it the only solution for an online application.
\end{abstract}


\section{Introduction}
\label{sec:introduction}

\IEEEPARstart{C}{ONVENTIONAL} power electronics technologies for energy conversion, such as two-level voltage source inverters and traditional hard-switched DC\,/\,DC topologies, still possess a high market penetration in industry and automation, but become inadequate for applications requiring higher voltage ratings, wide conversion ratios, galvanic isolation or improved output quality~\cite{wu2017high, perez2021modular,marquardt2018modular,hosseini2023energy,chen2020unbalanced,shang2021fast}. The inability of power semiconductors to withstand voltages beyond a few kilovolts, combined with the poor harmonic characteristics of two-level outputs, does not meet technical standards in high-voltage power conversion~\cite{rodriguez2002multilevel, leon2017multilevel}.

As an alternative, multilevel converters preferably employ a modular and scalable architecture with features such as the elimination of passive filters, improved reliability and redundancy, high power quality and density, the use of low-voltage semiconductors, and the reduction of common-mode voltage as well as electromagnetic interference~\cite{wu2017high, rodriguez2002multilevel, perez2021modular, marquardt2018modular}. These characteristics enable advances across applications including medium-voltage motor drives~\cite{li2020hybrid, kumar2019balanced, hagiwara2012startup, okazaki2014research}, offshore wind energy \cite{chen2025analysis}, high-voltage direct-current (HVDC) transmission~\cite{flourentzou2009overview, ansari2020mmc, sun2022beyond, li2018operation, sharifabadi2016design}, electric vehicles~\cite{goetz2015modular,kacetl2022ageing,tashakor2023generic,leon2017multilevel,kacetl2022bandwidth,8529274,Chuang2019PV}, and unified power quality conditioners~\cite{cheung2018transformerless}.

Considering the requirements of an application, there are a variety of modulation techniques, including carrier-based modulation ~\cite{hossain2025advanced}, space-vector modulation \cite{du2025analysis}, selective harmonic elimination \cite{wang2025selective}, nearest-level modulation ~\cite{wang2021flexible}, and various hybrid techniques that integrate several strategies ~\cite{10436634}. Carrier-based modulation methods, in particular, are often preferred due to their simplicity, ease of implementation, and suitability for applications in both low- and medium-voltage ranges.
Carrier-based methods assign a signal to each module or module interconnection (usually in the form of a sawtooth or triangle), where these signals can be distributed vertically (level-shifted-carrier PWM or LSC-PWM), horizontally (phase-shifted-carrier PWM or PSC-PWM) \cite{9275384}, or as a mix of both. These carriers then serve to generate the PWM switching signals for the corresponding modules or interconnections. Phase-shifted carriers are often preferred in high-power low-voltage applications with few modules \cite{kacetl2022ageing,tashakor2023generic,goetz2016sensorless,holmes2003pwm}. Under ideal conditions with balanced power modules and modulation indices, PSC-PWM can offer excellent performance. However, in real-world applications, such as photovoltaic battery systems \cite{8576676}, battery energy storage \cite{hannan2021battery}, but also grid-connected converters \cite{boscaino2024grid}, existing PSC concepts have suboptimal performance due to unbalanced operating conditions \cite{zhang2025topology,8000650,8341622}.

An increase of the number of modules offers a significant improvement in the current ripple. For example, in an interleaved multilevel DC\,/\,DC converter with \(N\) modules, the current ripple scales up to \(1/N^{2}\). However, current ripple remains a concern, as filters are generally designed for a specific current and voltage ripple profile \cite{majeed2018multiple,montesinos2012design,chen2022current}. Moreover, it is necessary to mitigate the unwanted harmonics of the output voltage since they contribute to both electromagnetic interference as well as losses. However,  the impact of different harmonic orders is in general not equal. For this reason, the weighted total harmonic distortion (WTHD) is considered as an important performance metric which assigns weights to harmonic orders according to their order of importance. Here, we use the order of the harmonics as their respective weights so that higher harmonics have a lower impact in the optimization problem.

Existing methods in the literature for optimizing the pulse patterns or the modulation techniques include software-based approaches that involve the adjustment of phase shift angles of carriers in PSC-PWM to improve the output voltage waveform. This method exploits PSC-PWM advantages, even under nonideal conditions. Jiao et al.\! offer a method that achieves real-time ripple mitigation with a phase-shifted PI controller for each module to reduce harmonics, but reduced harmonics come at the expense of additional controllers whose number scales with the modules~\cite{jiao2021closed}. An alternative approach relies on a harmonic model to determine optimal phase-shift angles~\cite{marquez2017variable,marquez2019generalized,monopoli2018improved,alcaide2021variable,liu2021derivation,an2023selective}. Yet, its complexity grows significantly with the number of modules. Marquez et al.\! suggest a varying phase shift to mitigate harmonics in a three-cell cascaded half-bridge converter~\cite{marquez2017variable}. Monopoli et al.\! take a similar approach and investigate various harmonic elimination schemes suitable for unbalanced operating conditions, but all of them are only applicable to three modules~\cite{monopoli2018improved}. Subsequent studies have attempted to extend these methods to a larger number of power modules. Alcaide et al., for instance, present an approach applicable to cascaded half-bridges with an increased module count~\cite{alcaide2021variable}. However, their solution has a high computational burden and cannot perform well when some of the modules are not operational.

Some other studies reduce computational time by simplifying the mathematical calculations; still, as the number of modules increases, computational complexity remains a major challenge. Some methods speed up computations and prevent the algorithm from getting stuck in local minima. Liu et al.\! suggest a phasor-diagram-based method to eliminate selective harmonics~\cite{liu2021derivation}. In the same way, An et al.\! present a method for full-range harmonic suppression that uses selective virtual synthetic vectors~\cite{an2023selective}. Despite advances in optimization techniques, current methods still require significant computation time and resources to determine optimal phase-shifting angles, relying on exhaustive searches or lookup tables. Even with improvements in data storage and processing, these methods struggle to adapt to complex real-world conditions, and exceeding time limits can delay computations.

On the other hand, conventionally AI and machine learning are limited for embedded online control in power electronics, mainly due to the computation and memory burden. 
Nevertheless, there is a growing trend in AI-based approaches in control-loops with smaller dynamics or when slower inferences would not lead to critical issues \cite{rajamony2022artificial,zhao2020overview,meshram2024advancements}. 

With presently available control methods and processing power, real-time optimization of optimal phase-shift angles based on system operating points is impractical. As a solution, we propose a supervised learning method with a simple, small neural network trained on a number of modules and modulation indices as well as the corresponding optimum carrier phase-shifts. The training data are derived offline through optimization with a genetic algorithm (GA), and no numerical optimization is necessary during the operation of the system. This method requires only one offline training stage, regardless of the number of modules, and can be adapted to fault conditions. In consequence, the method is applicable to systems with any number of modules. The proposed method is 100,000 to 500,000 times faster than online evolutionary or grid-search or heuristic optimization, such as GA, and requires far less memory than look-up table techniques \cite{Wang2019PV}. Moreover, the proposed method allows efficient reuse of the trained model without retraining, which reduces the computational effort and training time but still maintains consistent performance across varying system sizes. In fact, a neural network trained for a system with \textit{N} modules can be readily applied to systems with a total number of modules that is an integer multiple of \textit{N}.

\begin{figure}[!ht]  
    \centering
    \includegraphics[scale=0.5]{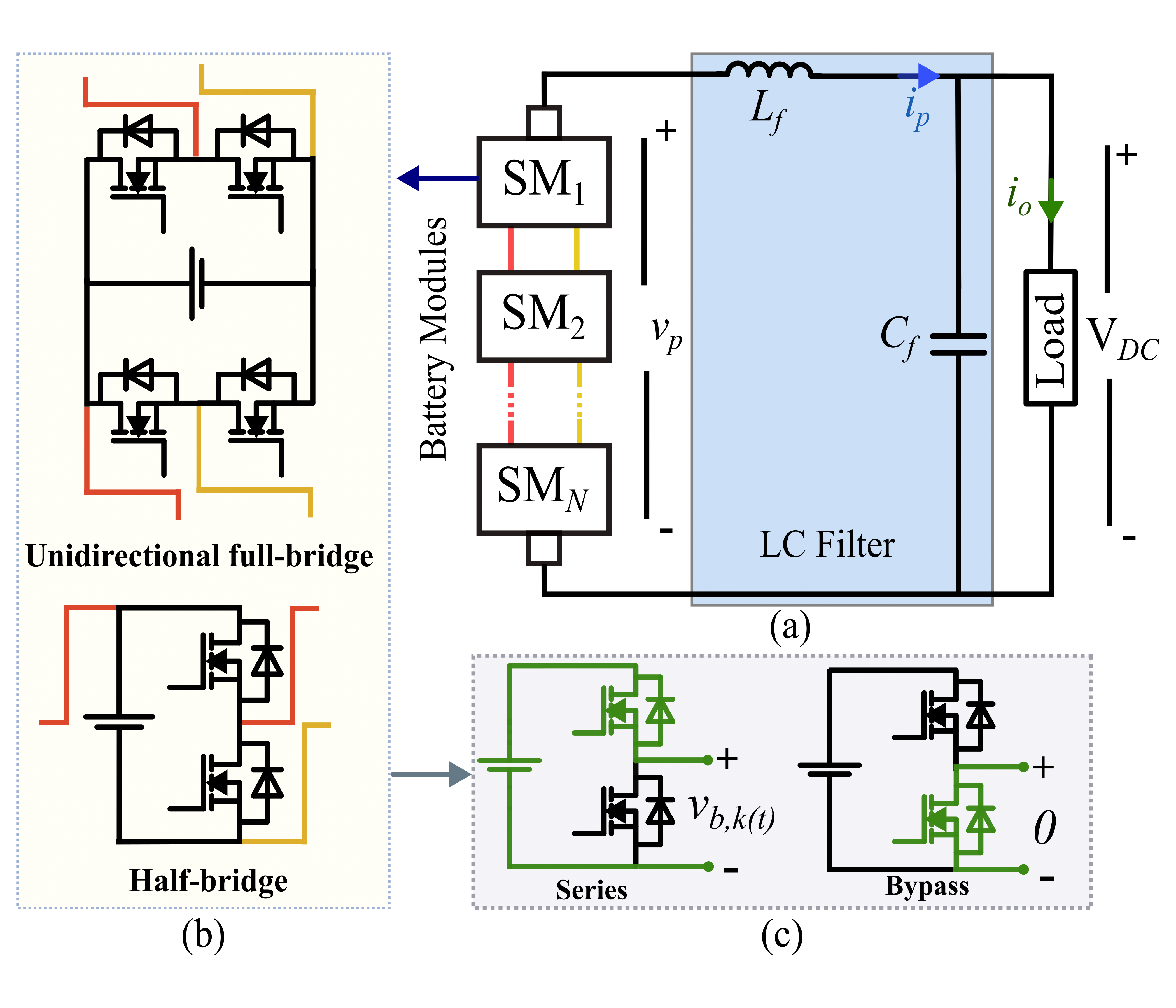}
    \caption{\footnotesize Generic reconfigurable battery system: (a) topology; (b) module type; (c) module operating modes.}
    \label{fig:1}
\end{figure}
\section {Applicable Modular Topologies: Reconfigurable Batteries as a Case Study}
This study mainly uses a battery string as the model circuit, as the string is the basic topology for more complicated module assemblies, such as star, delta, or the Marquardt arrangement. Thus, the proposed framework can be applied to all assemblies, such as modular multilevel converters, cascaded bridges, and other variants of reconfigurable batteries that contain individual strings. We discuss this in Section~\ref{sec:IV} in more detail.

\subsection {Circuit Description}
The generic structure of reconfigurable batteries with a single string (see Figure~\ref{fig:1}~(a)) consists of \( N \) cascaded battery-integrated power modules with parallel mode, which connect to the next stage of power conversion (here simply named the load) through a low-pass filter (\( L_f \) and \( C_f \)). In reconfigurable batteries, the unidirectional arrangement, such as half-bridge and full-bridge (see Figure~\ref{fig:1}~(b)), is often also preferred over more advanced ones, such as the series\,/\,parallel module arrangement to simplify operation at the cost of reduced performance, flexibility, and battery utilization. For the sake of understanding, we present the entire concept for half-bridge power modules as the simplest topology. Nevertheless, the concept can be easily extended to more complex concepts, as discussed later on. The half-bridge module operates in series or bypass (see Figure~\ref{fig:1} (c)) and the formed voltage for the entire battery string follows
\begin{equation}
v_p(t) = \sum_{k=1}^{N} s_k(t) v_{bk}(t).
\label{eq:1}
\end{equation}
In \eqref{eq:1}, \( v_{bk}(t) \) represents the battery terminal voltage of the \(k\)-th power module. In addition, \( s_k(t) \in \{0, 1\} \) denotes the corresponding state.
As a result of PSC-PWM, the string voltage will be a pulsating step-shaped variable voltage (see Fig. 2). 
In addition, the low-pass filter converts this modulated voltage \( v_p(t) \) to a refined voltage across the load. The ideally filtered voltage across the load using is
\begin{equation}
v_o(t) = \sum_{k=1}^{N} m_k V_{oc,k},
\label{eq:2}
\end{equation}
where \( m_k \) is the modulation index and \( V_{\textrm{oc},k} \) is the average voltage of Module $k$. Furthermore, \( i_o(t) \) stands for the load current and \( i_p(t) \) represents the current of the battery pack. However, as Section II-B explains, no filter is really ideal.
\subsection {PSC-PWM and Problem Description}
\begin{figure}[!ht]  
    \centering
    \includegraphics[scale=1]{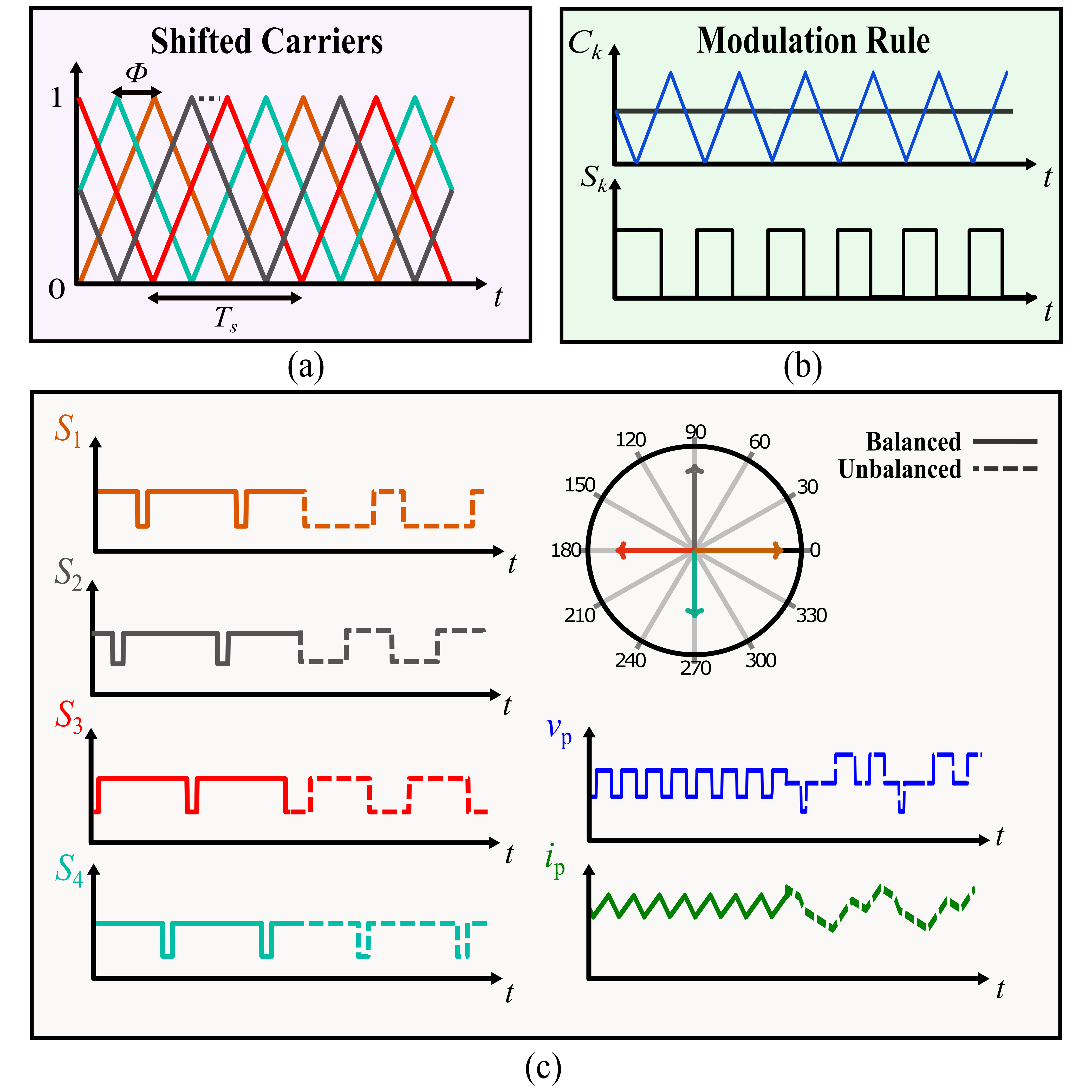}
    \caption{PSC-PWM features and performance: (a) shifted carriers; (b) modulation rule; (c) performance under balanced and unbalanced operating conditions for a case study of four modules.}
    \label{fig:2}
\end{figure}

Conventional PSC-PWM uses $N$ uniform carriers with the same frequency, evenly distributed within one switching cycle. The phase angle of the $k$-th carrier is calculated per $\varphi_k = \frac{2k\pi}{N}$, and the phase shift between each pair of adjacent carriers is $\varphi = \frac{2\pi}{N}$ (Figure~\ref{fig:2}(a)). According to the modulation rule, $s_k$ is set to one when the reference exceeds the carrier and to zero when the opposite is true (Figure~\ref{fig:2}(b)).
In case of evenly-spread carriers and balanced modulation indices and module voltages, PSC-PWM offers the optimum output quality. However, when the modulation indices differ, the voltage of the battery pack becomes distorted. For example, in a reconfigurable battery system with four modules (Figure~\ref{fig:2}(c)), balanced modules with identical modulation indices generate evenly distributed pulses, which form a two-level battery pack voltage with minimal ripple current. In contrast, when modulation indices vary (e.g., due to battery charge balancing), the differing pulse widths of the module outputs distort the aggregated string voltage and increase current fluctuations (Figure~\ref{fig:2}(c)). The evenly spread phase-shift angles of carriers per conventional PSC-PWM assume a balanced system and control. Consequently, under non-ideal conditions, the voltage of the deteriorated battery pack leads to an increase in current ripple and WTHD. Hence, we intend to identify phase angles that reduce both current ripple and WTHD as a key to enhance performance.
\subsection{Current Ripple and Harmonic Analysis}
When the system is unbalanced, it is advantageous to employ the frequency domain model, as it offers a deeper insight into how the \( n \)-th harmonic order affects the current ripple. Under the assumption that the module voltages are well balanced, we apply fourier-transform to the switching signal of the \( k \)-th module, which includes a phase shift \( \varphi_k \) at each sampling period \( T_s \) as
\begin{equation}
    s_k(t) = \frac{a_{0k}}{2} + \sum_{n=1}^{\infty} \left( a_{nk} \cos(n\omega t + \varphi_k) + b_{nk} \sin(n\omega t + \varphi_k) \right)\!,
    \label{eq:3}
\end{equation}
where \( \omega = \frac{2\pi}{T_s} \). Furthermore, \( a_{0k} \), \( a_{nk} \), and \( b_{nk} \) are the coefficients given by
\begin{equation}
    \begin{cases}
        a_{0k} = 2m_k, \\
        a_{nk} = \frac{2}{n\pi} \sin(n\pi m_k), \\
        b_{nk} = 0.
    \end{cases}
    \label{eq:4}
\end{equation}
Substituting Equations (\ref{eq:2}) and (\ref{eq:3}) into the expression for \( v_p(t) \), we obtain
\begin{align}
    v_p(t) &= v_o(t) + \sum_{k=1}^{N} \sum_{n=1}^{\infty} \frac{2 V_{oc}}{n \pi} \sin(n \pi m_k) \cos(n \omega t + \varphi_k),
    \label{eq:5}
\end{align}
where \( V_{oc} \) is the average value of the modules, which is independent of harmonic order \( n \). The remaining pulsating voltage across the inductor is
\begin{align}
    L \frac{\textrm{d}i_p}{\textrm{d}t} = \sum_{k=1}^{N} \sum_{n=1}^{\infty} \frac{2 V_\textrm{oc}}{n \pi} \sin(n \pi m_k) \cos(n \omega t + \varphi_k),
    \label{eq:6}
\end{align}
which causes current variation per
\begin{equation}
    \Delta i_p(t) = \sum_{k=1}^{N} \sum_{n=1}^{\infty} \left[ \frac{2}{L \omega n^2 \pi} \sin(n \pi m_k) \sin(n \omega t + n \varphi_k) \right] + i_0.
    \label{eq:7}
\end{equation}
The ripple caused by the \( n \)-th harmonic is given by
\begin{equation}
\Delta i_{p,n}(t) = 
    \begin{aligned}
        &\cos(n\omega t) 
        \sum_{k=1}^{N} \frac{2V_\textrm{oc}}{L \omega n^2 \pi} \sin(n \pi m_k) \sin(n \varphi_k) \\
        &+ \sin(n\omega t) 
        \sum_{k=1}^{N} \frac{2V_\textrm{oc}}{L \omega n^2 \pi} \sin(n \pi m_k) \cos(n \varphi_k),
    \end{aligned}
\label{eq:8}
\end{equation}
where the amplitude of the \( n \)-th harmonic is
\begin{equation}
A_n = \sqrt{
    \begin{aligned}
        &\left(
            \sum_{k=1}^{N} \frac{2V_\textrm{oc}}{L \omega n^2 \pi} 
            \sin(n \pi m_k) \sin(n \varphi_k)
        \right)^2 \\
        &+ \left(
            \sum_{k=1}^{N} \frac{2V_\textrm{oc}}{L \omega n^2 \pi} 
            \sin(n \pi m_k) \cos(n \varphi_k)
        \right)^2\!\!\!.
    \end{aligned}
}
\end{equation}
Although reducing THD generally decreases current ripple, it does not necessarily lead to effective suppression of low-order harmonics, which significantly influence system performance. For this reason, the study considers WTHD, which accounts for both the amplitude and frequency of each harmonic order. WTHD is calculated as
\begin{equation}
\text{WTHD}_\textrm{dc} = \frac{\sqrt{ \sum_{n=1}^{\infty} w_n \cdot A_n^2 }}{V_\textrm{dc}}, 
\label{eq:10}
\end{equation}
where \( w_n \) is the weight factor.
\section{Proposed Framework}
\subsection{Neural-Network Modulation and Scheduling Framework}
\begin{figure}[!ht]  
    \centering
    \includegraphics[scale=1]{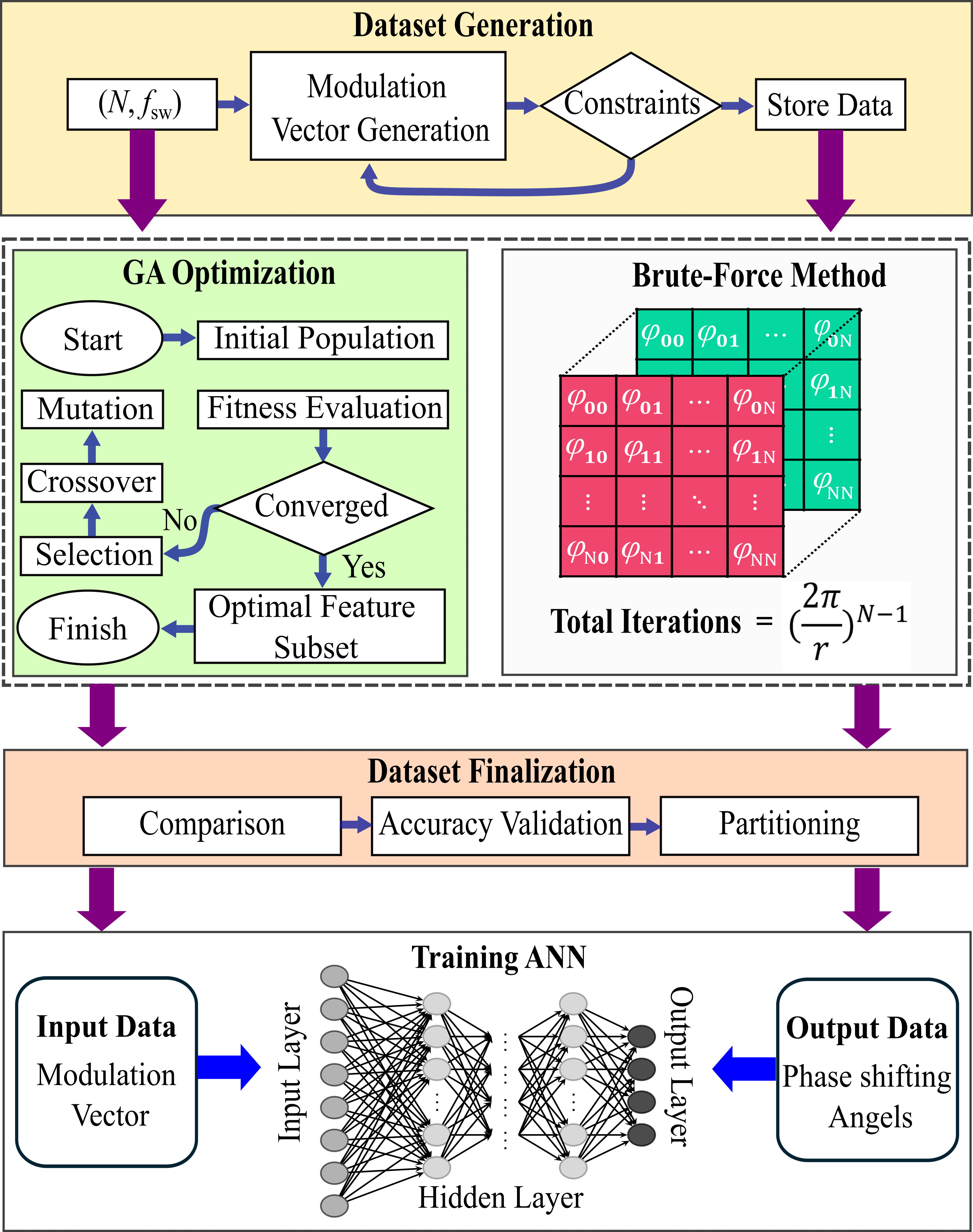}
    \caption{Workflow of the proposed framework.}
    \label{fig:3}
\end{figure}

The core scheduling algorithm is implemented as a supervised learning model in the form of a feedforward neural network. Figure~\ref{fig:3} illustrates the proposed method framework. The neural network is trained to predict the optimal phase-shift angles directly from the modulation vector, hence avoiding the pitfall of the online grid-search or heuristic optimization in the state of the art. We employ a genetic algorithm offline to build a dataset by computing the optimal phase shifts for a wide range of operating conditions. This GA-based labeled dataset generation step is performed only once and is not part of the real-time control process. After training, the neural network fully replaces the GA, which enables real-time prediction of phase-shift angles with negligible computational cost compared to iterative optimization.
\begin{figure}[!ht]  
    \centering
    \includegraphics[scale=1]{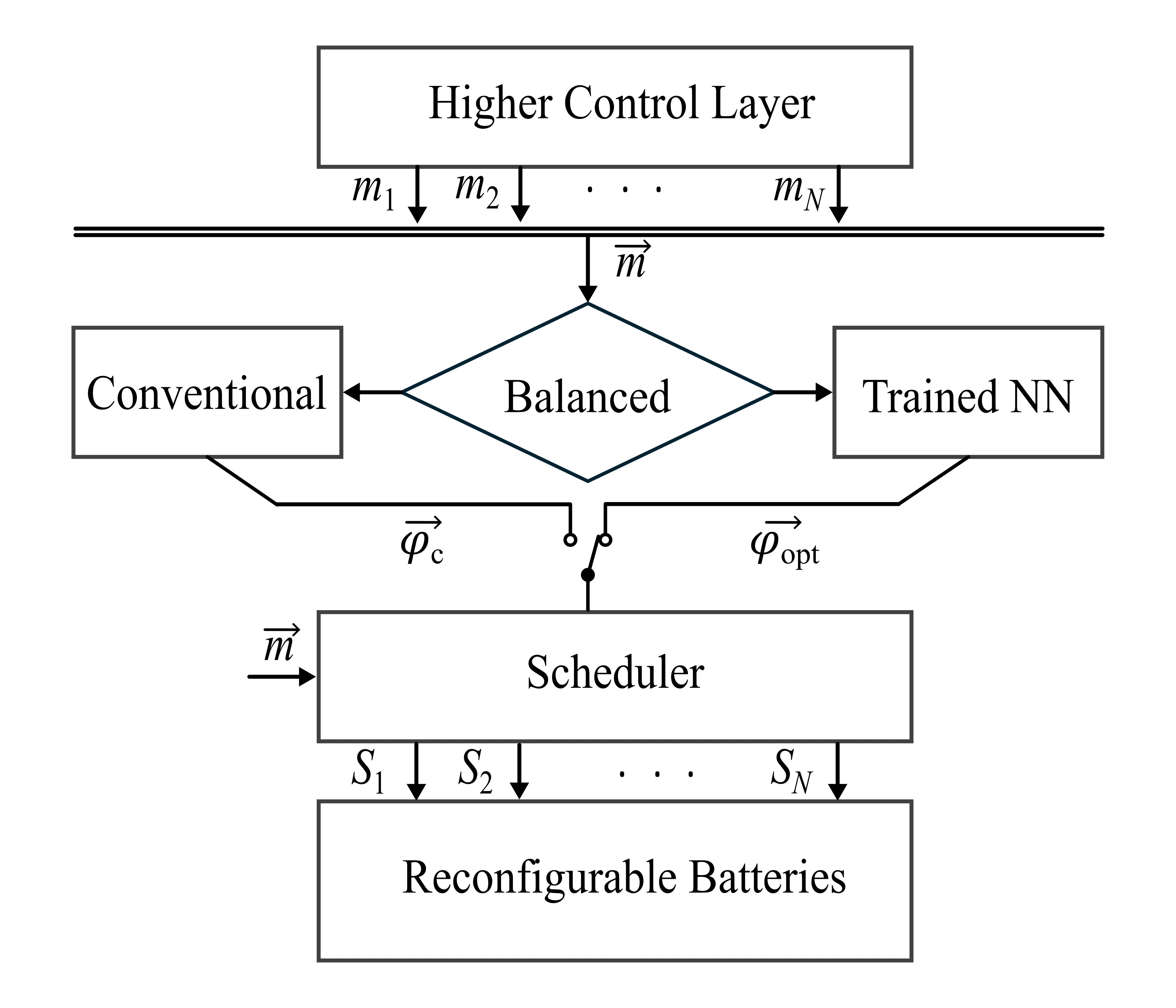}
    \caption{The proposed NN integration with control scheme.}
    \label{fig:4}
\end{figure}
\subsubsection{Sample Generation}
The proposed method uses the derived models for the output voltage and current---verified by experiments---to efficiently generate the necessary data set. The framework begins by defining the number of modules \( N \) and the switching frequency \( f_{\text{sw}} \), followed by the generation of a set of random modulation vectors \( \boldsymbol{m} = [m_1, m_2, \ldots, m_N] \), with the resolution of 0.02, each bounded within a feasible range as
\begin{equation}
    0 \leq m_i \leq 1, \quad \forall i \in \{1, 2, \ldots, N\}.
    \label{eq:modulation_range}
\end{equation}
Furthermore, the complementary modulation vector is defined as 
\begin{equation}
    \mathbf{\textit{m}}' = [1 - m_1, 1 - m_2, \dots, 1 - m_N].
    \label{eq:complement}
\end{equation}
Although the modulation indices range between 0 and 1,  the optimum pulse patterns for $\mathbf{\textit{m}}$   and $\mathbf{\textit{m}}'$ are identical. Therefore, we only generate random patterns between 0 and 0.5 to reduce the training complexity. 
Subsequently, we sort the modulation vector elements in an increasing order to prevent non-identical patterns. Moreover, according to the randomly generated datasets the peak deviation among the ordered modulation vectors remove the non-unique samples as
\begin{equation}
    \max(|\mathbf{\textit{m}}_\text{sorted} - \mathbf{\textit{m}}'_\text{sorted}|) \geq \epsilon,
    \label{eq:tolerance_check}
\end{equation}
where \(\epsilon > 0\) is a tolerance threshold to account for floating-point inaccuracies as well as account for the resolution of the pulses that can be generated in real world. Then, we identify and remove modulation vectors with duplicate or nearly identical elements to ensure an unbiased training data set. It is important to note that we include extreme modulation cases where any index is equal to 0 or 1, which translates the operating conditions that some of the modules are not operational or need to be bypassed.

\subsubsection{Cost Function Design and Feature Selection}
The cost function design procedure considers current ripple and WTHD as control objectives. We define the cost function with the model derived in Section II, which was previously verified through experiments, as
\begin{equation}
J({\varphi}, {m}) = w_1 \cdot \Delta i_p({\varphi}, {m}) + w_2 \cdot \text{WTHD}_\textrm{dc}({\varphi},{m}),
    \label{eq:subset_sum}
\end{equation}
where \({\varphi} = [\varphi_1, \varphi_2, \ldots, \varphi_{N-1}] \) is the vector of phase shift angles that should be adjusted to minimize the cost function. Phase shifts are constrained by periodicity as
\begin{equation}
\varphi_i \in [0, 2\pi), \quad \forall i \in \{1, 2, \ldots, N_c-1\}.
    \label{eq:subset_sum1}
\end{equation}
Furthermore, \( w_1 \) and \( w_2 \) are respectively weighting factors for current ripple and WTHD, which can be arbitrarily set. Here, we set both to one to give equal priority. Given that WTHD is unitless and the current ripple has a unit, it is essential to normalize them. For normalization, we have selected a phase-shift vector of zero, which means that battery modules function without phase difference. The initial population of phase-shift angles is discretized with a step size of \( \pi/5 \) to ensure an efficient search resolution. The optimization process runs for up to 300 generations with a convergence tolerance of \(10^{-6}\). 

\subsection{Comparison Between GA Optimization and Exhaustive Search}
We further evaluate the performance of genetic optimization by comparing the results obtained with an exhaustive permutation-combination method. The brute-force search method discretizes the phase shifts with an angle resolution of \( r\). This exhaustive search employs \( N-1 \) nested loops on phase-shift vectors. Generally, the total number of possible phase shift configurations is calculated as
\begin{equation}
\text{Total Iterations} = \left(\frac{2\pi}{r}\right)^{\!\!N-1}\!\!\!\!\!\!\!\!\!\!,
\end{equation}
where \( \frac{2\pi}{r} \) represents the possible values per phase for phase shifts. The performance threshold metric is defined as
\begin{equation}
\text{Threshold} = \frac{C_{\text{GA}} - C_{\text{perm}}}{C_{\text{perm}}} \times 100\%,
\end{equation}
where \( C_{\text{GA}} \) is the value of the cost function obtained by GA and \( C_{\text{perm}} \) is the optimal cost function of the exhaustive search. If the threshold is low, it means that the genetic algorithm is very close to the optimal solution with brute force. If the threshold is high, it means that the genetic algorithm finds better phase-shift configurations. In the end, the better one of the two alternatives are used for the next stage.

\subsection{Neural Network Training}
As the problem of optimizing phase angles is static, this research involves a feed-forward neural network (Figure~\ref{fig:3}). Table \ref{tab:ann_hyperparameters} lists the finalized optimized neural network features with its hyperparameters. The split ratio between training and testing is 85\% to 15\%. Moreover, Figure~\ref{fig:4} demonstrates the incorporation of the proposed neural network into the conventional control scheme. When the modulation indices differ, the method applies the predicted phase shifts from the neural network, whereas it uses traditional evenly-distributed phase shifts when the modulation indices are equal. It is important to note that the proposed approach is not responsible for balancing the modules. Normally, outer control layers perform the balancing routine. In other words, higher-level controllers (voltage and current controller as well as the balancing and safety layers) determine the needed modulation vector, and the proposed approach provides the most optimum solution given the modulation vector.

\begin{table}[t]
\centering
\caption{Neural Network Hyperparameters and Architecture Details}
\label{tab:ann_hyperparameters}
\begin{tabular}{c | c}
\hline \hline
\textbf{Hyperparameter} & \textbf{Value} \\ \hline
\multicolumn{2}{c}{\textbf{Network Characteristics}} \\ \hline
Network Type & Feed-forward neural network \\ 
Number of Layers & 7 \\ 
Neurons per Layer & \{729, 243, 81, 27, 9, 3\} \\ 
Activation Function & \(\tanh\) (hidden layers), linear (output layer) \\ 
\hline
\multicolumn{2}{c}{\textbf{Training Parameters}} \\ \hline
Optimizer & Adam \\ 
Loss Function & Mean absolute error (MAE) \\ 
Initial Learning Rate & \(1 \times 10^{-3}\) \\ 
Regularization & L2 regularization (\(\lambda = 1 \times 10^{-4}\)) \\ 
Batch Size & 512 \\ 
Maximum Epochs & 2000 \\ 
Validation Split & 5\% \\ 
Validation Frequency & Every 500 iterations \\ 
Early Stopping Patience & 50 epochs \\ 
Learning Rate Schedule & Piecewise decay \\ 
Random Seed & 42 \\ 
Shuffling Strategy & Every epoch \\ 
\hline
\end{tabular}
\end{table}

\begin{figure}[!ht]  
    \centering
    \includegraphics[scale=1 ]{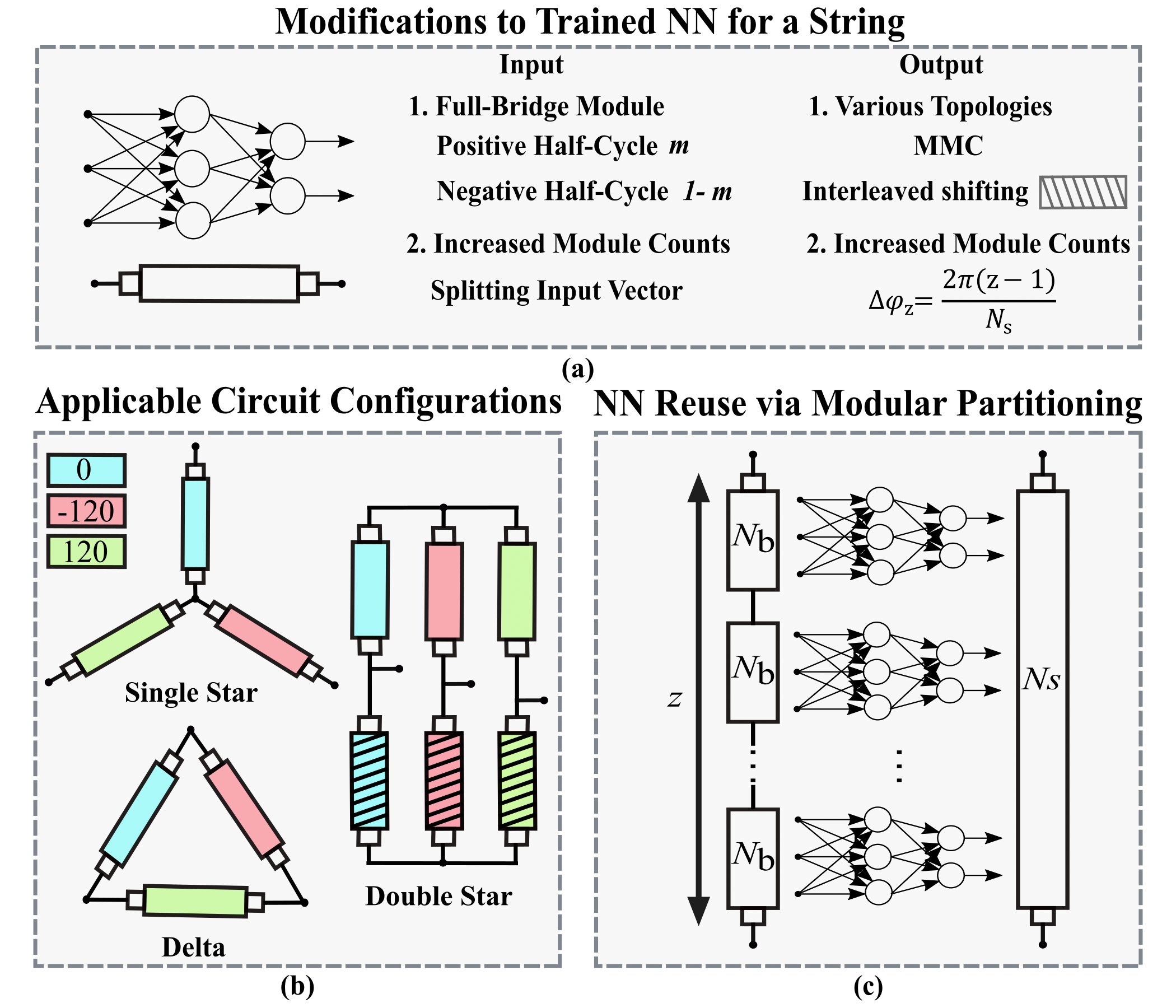}
    \caption{The proposed method expandability.}
    \label{fig:5}
\end{figure}
\section{Expandability to Other Topologies, Module Counts, and Carrier Waveforms}
\label{sec:IV}
This section discusses the applicability of the proposed method to other converter topologies employing PSC-PWM, its scalability with respect to the number of modules, and the impact of carrier waveform shape. The main idea is to reuse the proposed method for a single string and expand it through modifications at modulation vector and phase shift angles. Figure~\ref{fig:5} visualizes the different adjustments required to apply to the modulation vector and predicted phase angles.

\subsection{Applicable Circuit Configurations}
The proposed framework determines the optimal distribution of carrier phase displacements as a function of the modulation index. The phase-shift vector is valid for both positive and negative polarities as it represents carrier timing rather than absolute voltage. Consequently, the neural network can serve for the negative polarity half-cycle by feeding a mirrored modulation signal, \( m'(t)=1-m(t) \), instead of \(m(t)\). This alteration inverts the comparison logic, which calculates the correct switching sequence for negative voltage polarity and still preserves the network's learnt temporal associations. 

The proposed framework can be extended to more intricate converter designs, such as delta, single-star, and double-star (MMC), without the need to retrain the neural model. In three-phase designs, the reference signals are phase-shifted to match the balanced three-phase system. However, for MMC designs, the reference signals for the upper and lower arms differ, which suggests an additional interleaved phase-shifting method to effectively mitigate circulating currents.

\subsection{Neural Network Reuse via Modular Partitioning}
The complexity of the training process grows with increasing numbers of modules. Although the proposed method solves this growth by requiring only a single training session for a specific base configuration with $N_b$ modules, the trained network can then be generalized to larger systems, provided that the scaled number of modules $N_s$ is an integer multiple $z$ of the base configuration, i.e., $N_s = z \cdot N_b$.
First, we divide the modulation vector with length $N_s$ into $z$ sub-vectors with $N_b$ elements \cite{Divide}. Each sub-vector is processed by the neural network trained for $N_b$ modules, which yields the corresponding phase shift. For the first sub-vector ($z = 1$), the predicted phase shift is directly applied without any change. For subsequent sub-vectors ($z = 2, 3, \dots, N_s/N_b$), the predicted phase shift is incremented by an offset
\begin{equation}
\Delta\varphi_z = \frac{2\pi (z-1)}{N_s},
\end{equation}
which accounts for the spatial distribution of the modules within the larger system. This approach of modular reuse enables neural networks that have been trained for smaller setups (such as with $N_b = 3$, $4$, or $5$ modules) to be adapted for use in systems featuring a greater number of modules, without the need for retraining. Although the improvements achieved  typically fall short of those of a network trained specifically for $N_s$ modules, they still exceed the conventional approaches.

\subsection{Carrier Waveform Mapping: Triangle to Sawtooth}
In this study, we trained the neural network with a center-aligned (triangular) carrier. However, the optimum phase-shift angles $\phi_i^{\mathrm{tri}}$ obtained in this case can be translated to other common carrier shapes, such as edge-aligned (saw-tooth) carriers, without retraining. The mapping is based on the fact that the intersection point between the modulation signal and the carrier shifts by a modulation-index-dependent offset. The triangular carrier yields a symmetric switching instant, whereas the sawtooth carrier produces an asymmetric one. The two can be related as
\begin{equation}
\phi_i^{\mathrm{saw}} 
= \mathrm{wrap}_{2\pi} \!\left( \phi_i^{\mathrm{tri}} \; \pm \; \frac{\pi}{2} m_i \right)\!,
\label{eq:phi_mapping_carrier}
\end{equation}
where $m_i$ is the modulation index of the $i$-th module, and 
$\mathrm{wrap}_{2\pi}(\cdot)$ denotes modulo-$2\pi$ wrapping.

\section{Results and Discussion}
Our case study is a reconfigurable battery system with four modules, which assigns nonidentical modulation indices to battery modules with the same voltage level. We use simulations based on verified modules to assess the statistical behavior of the proposed solution with a large sample space. Moreover, the provided experimental validations confirm the feasibility of the proposed solution in an online setting. 

\subsection{Large-Scale Assessment}
\begin{figure}[!ht]  
\includegraphics[width=0.5\textwidth]{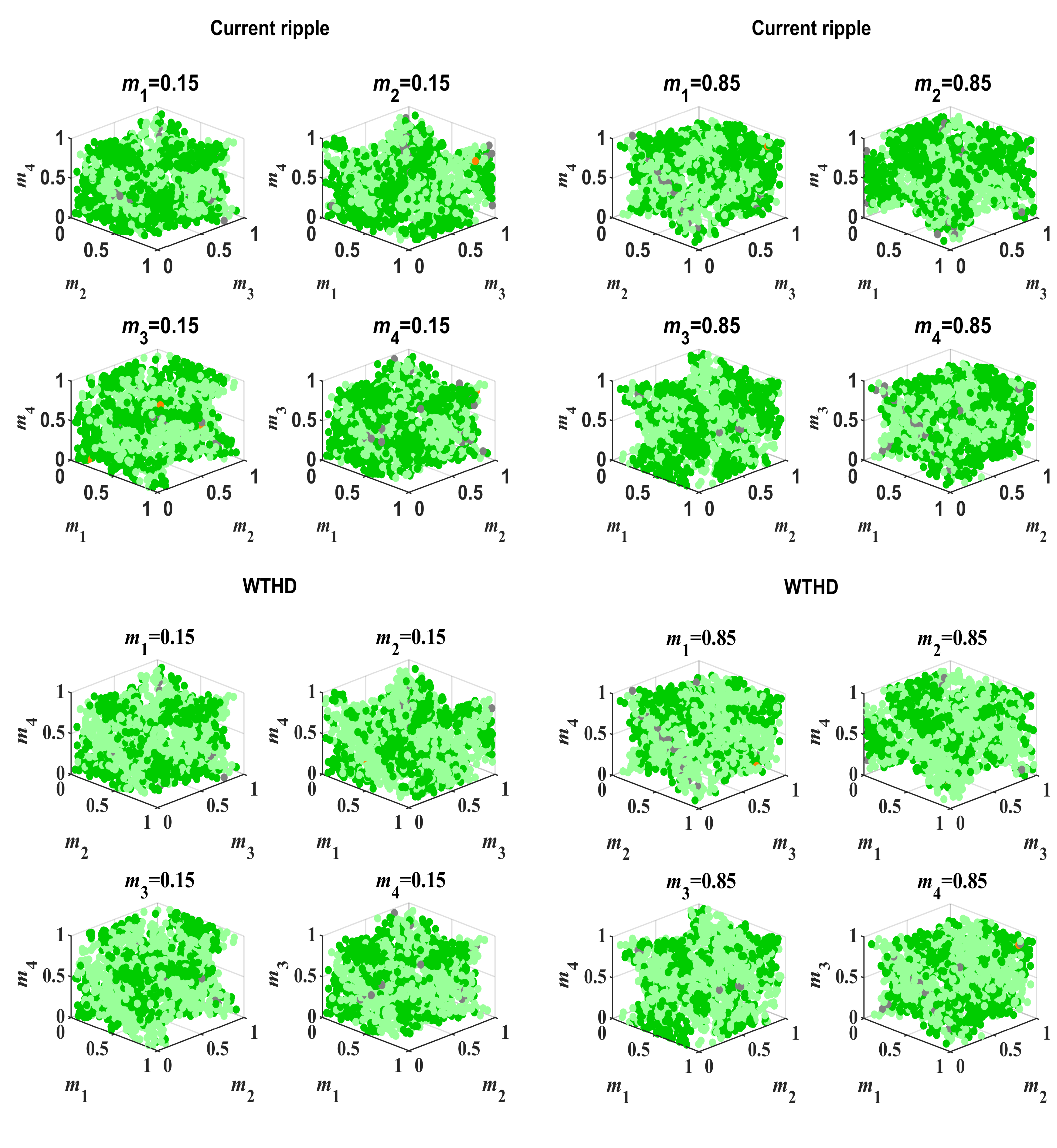}
    \caption{Large-scale performance assessment of the trained ANN.}
    \label{fig:6}
\end{figure}

The evaluation compares the predictions with traditional and genetic algorithm methods with respect to the resulted current ripple and WTHD. Scatter plots illustrate the results (Figure~\ref{fig:6}). Since a unified 4D representation is not possible, we fix one of the modulation indices in each plot and  vary the other three from 0 to 1. The results are color-coded for clarity: green indicates a performance within 1\% of the genetic algorithm or better than the traditional even carrier distribution, and gray represents results within 1\% of the conventional method.
The neural network demonstrates remarkable accuracy, where more than 96.88\% of the predictions match or exceed the genetic algorithm's performance. In particular, 499,040 cases (43.38\%) achieve results within 1\% of the genetic algorithm, and 615,460 cases (53.50\%) surpass the conventional approach. These results confirm the neural network's effectiveness in predicting the optimum phase-shift vector with reduced computational burden. The reported results are for the case of using 10000 random modulation vectors which is used for training process.

\begin{figure}[!ht]  
    \centering
    \includegraphics[scale=1]{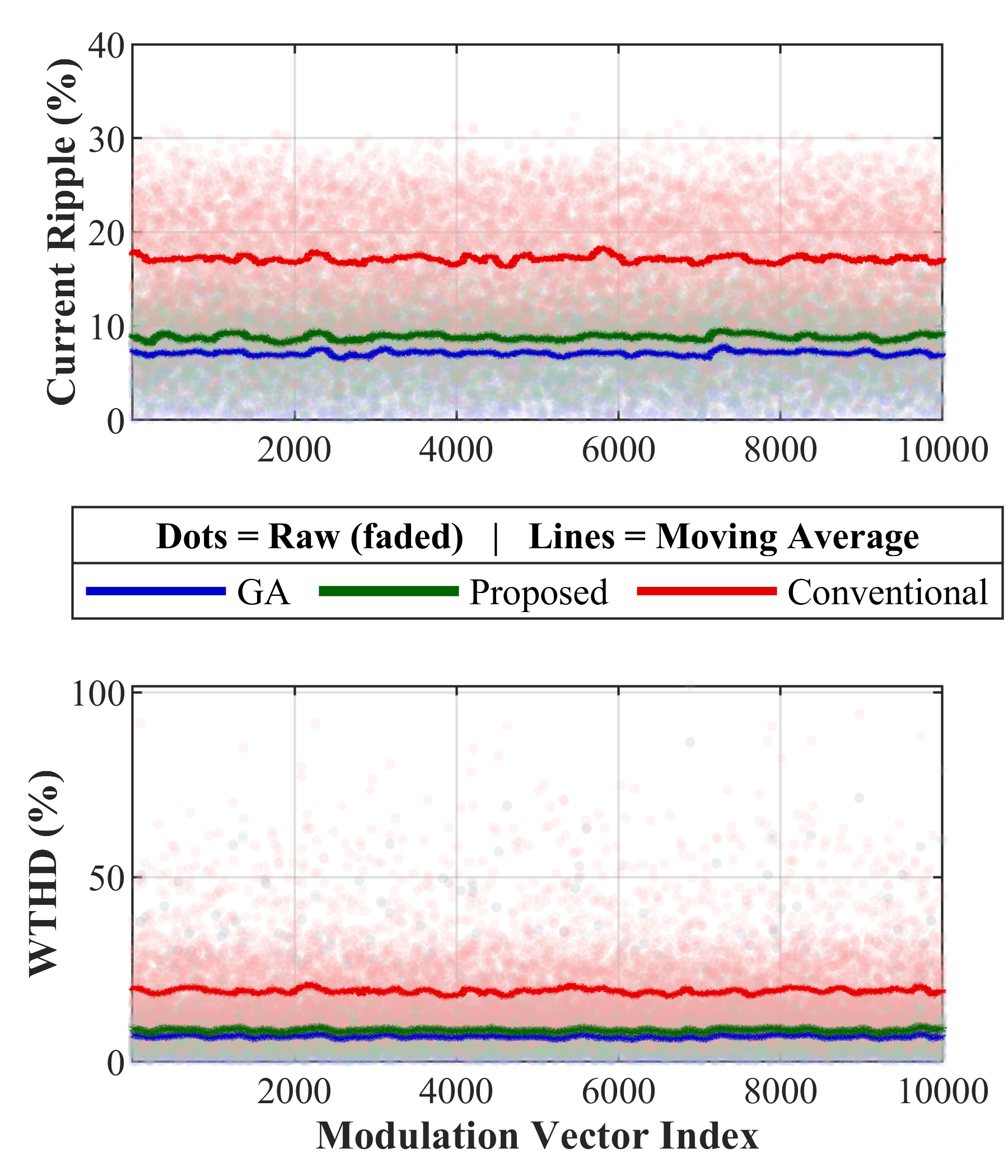}
    \caption{Performance comparison between the GA optimizer, proposed method, and Conventional method in terms of current ripple and WTHD.}
    \label{fig:7}
\end{figure}

\begin{figure}[!ht]  
    \centering
    \includegraphics[scale=1]{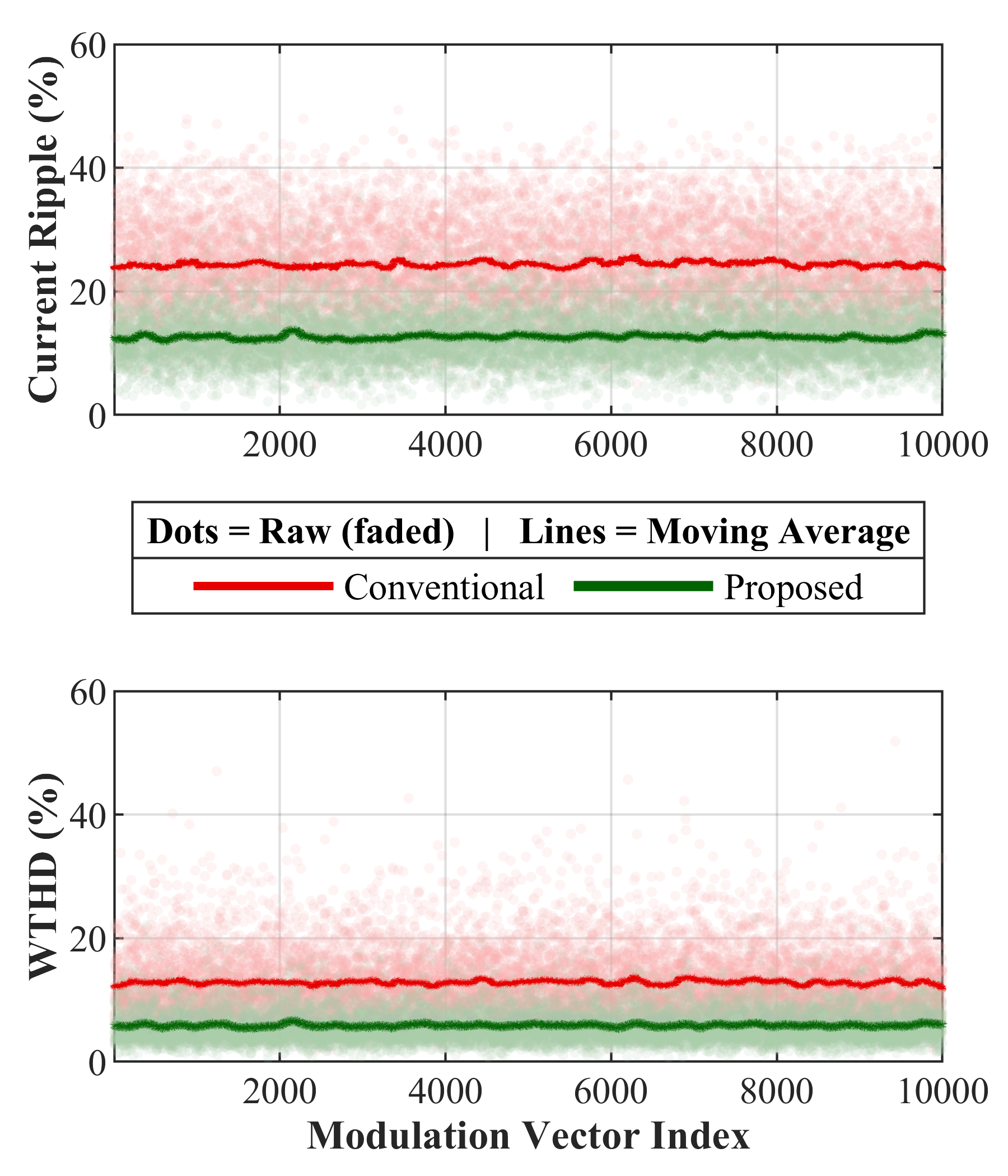}
    \caption{\footnotesize Performance comparison between the proposed neural network reuse method (via modular partitioning) and the conventional method in terms of current ripple and WTHD for the case of eight modules.}
    \label{fig:8}
\end{figure}

To further assess the performance of the proposed framework, we performed a comparative analysis with an additional set of 10,000 randomly chosen modulation vectors. The findings indicate that the proposed framework can closely replicate the accuracy of a GA optimizer. Figure~\ref{fig:7} illustrates the performance of the proposed method and the GA optimizer against traditional fixed phase shifts. On average, the proposed method reduces the current ripple and WTHD by 50\%.
Lastly, we present the approach as well as the results for extending or adapting an already developed model for a larger number of modules.
Figure~\ref{fig:8} compares the current ripple and WTHD obtained using the conventional method and the proposed modular partitioning approach for the case of $N_s = 8$ modules and $10{,}000$ random modulation vectors. On average, the proposed method achieves $40\%$ reduction in both current ripple and WTHD compared to the conventional method. Although the amount of reduction is not as significant as when the neural network is trained specifically for eight modules, the proposed approach still achieves high performance as it reaches approximately 90\% of the accuracy obtained for a single string with four modules.
\subsection{Experiments}
\begin{figure}[!ht]
    \centering
    \includegraphics[scale=1]{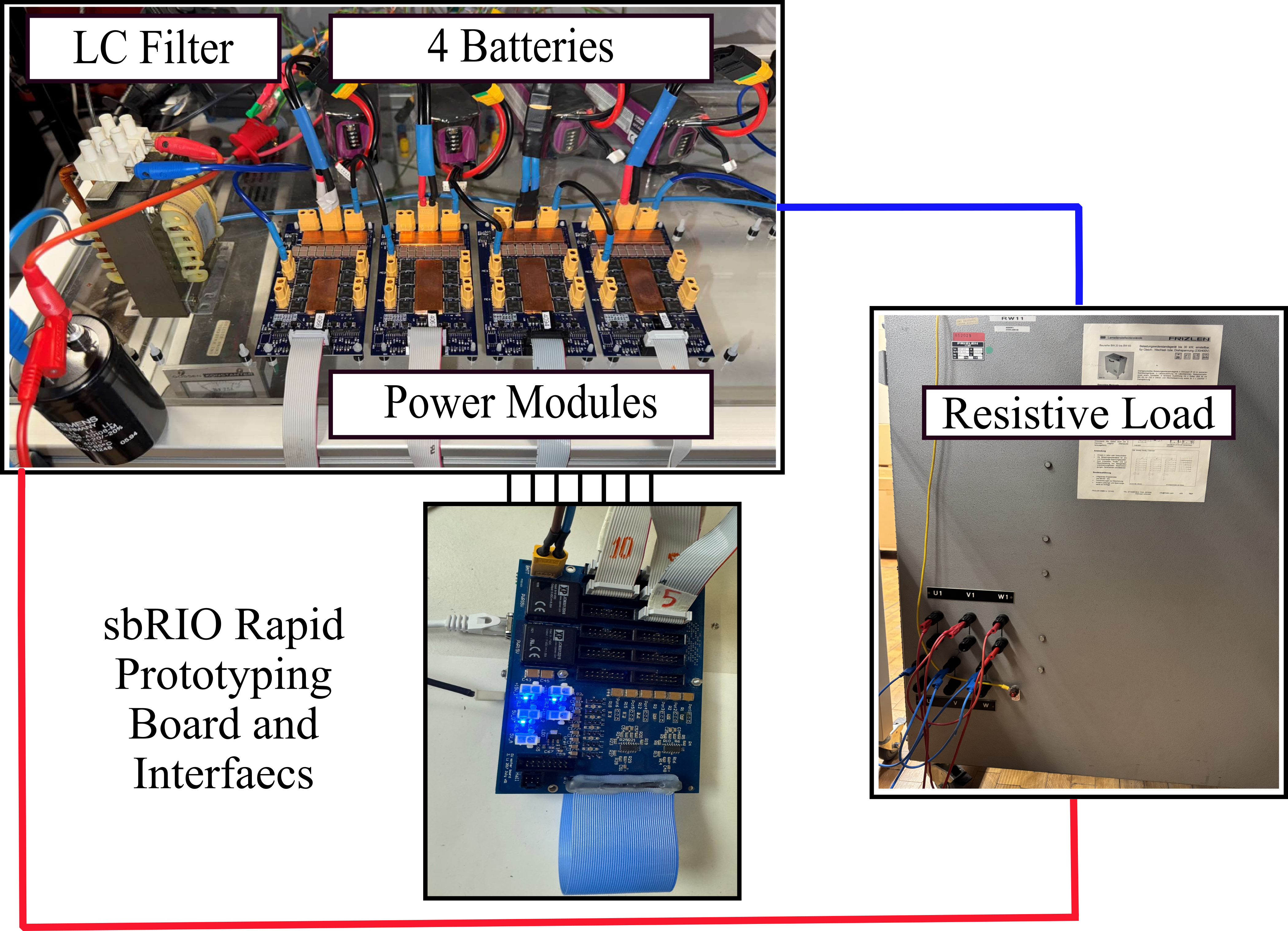}
    \caption{Reconfigurable-battery test bench.}
    \label{fig:9}
\end{figure}

 We implemented the simulated system in the experiments as well (Figure~\ref{fig:9} and Table~\ref{tab:ripple_wthd_comparison}). Each module consists of four half-bridge legs with high-current field-effect transistors (Infineon IPT015N10N5). Low-ESR ceramic capacitors form a 500 µF commutation capacitance in each module, parallel to a lithium-polymer battery in a 6s2p configuration with a nominal voltage of 22.2 V and a capacity of 5 Ah. A commercially available rapid-control prototyping board (sbRIO 9627 from National Instruments) implements the controller with a Zynq-7020 system on chip. Nevertheless, the final neural network size is small enough that it can be loaded into a low-cost micro-controller.

For a targeted and representative validation, we assess the method using four specific modulation vectors that cover various unbalanced case studies. We also provide a cross validation of these cases against simulation results. The module voltages can deviate due to tolerance, SoC, and small impedance, and the load conditions can also differ. Nevertheless, simulation and experimental results widely match with minimal deviation in the absolute values of ripple and WTHD. Furthermore, the comparative analysis shows that the optimized phase shifts in both simulation and experiments are optimal.

\begin{figure}[!ht]  
    \centering
    \includegraphics[scale=0.5]{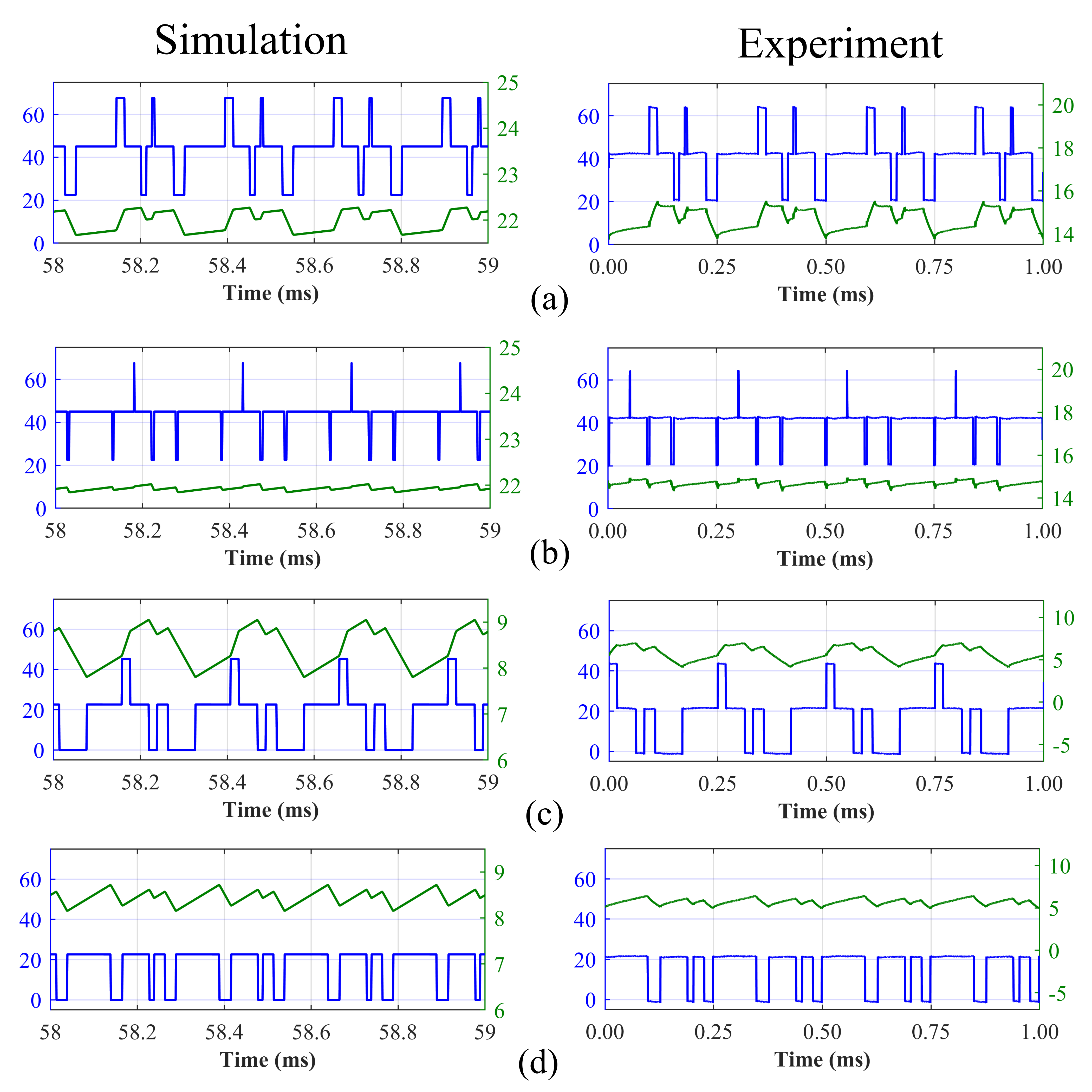}
    \caption{\footnotesize Battery pack output voltage and current. 
    First scenario: (a) conventional PSC-PWM; (b) proposed method.  
    Second scenario: (c) conventional PSC-PWM; (d) proposed method.}
    \label{fig:10}
\end{figure}

In the first scenario, where all modulation indices are below 0.5 (\(\mathbf{\textit{m}} = [0.15, 0.3, 0.4, 0.45]\)), conventional PSC-PWM causes mismatch pulses at the pack voltage, which results in increased current ripple and WTHD compared to balanced conditions (Figure~\ref{fig:10}(a)). The proposed method eliminates these mismatched pulses through optimal predicted phase shifts, which simultaneously reduce current ripple and WTHD  (Figure~\ref{fig:10}(b)). 

\begin{figure}[!ht]  
    \centering
    \includegraphics[scale=0.5]{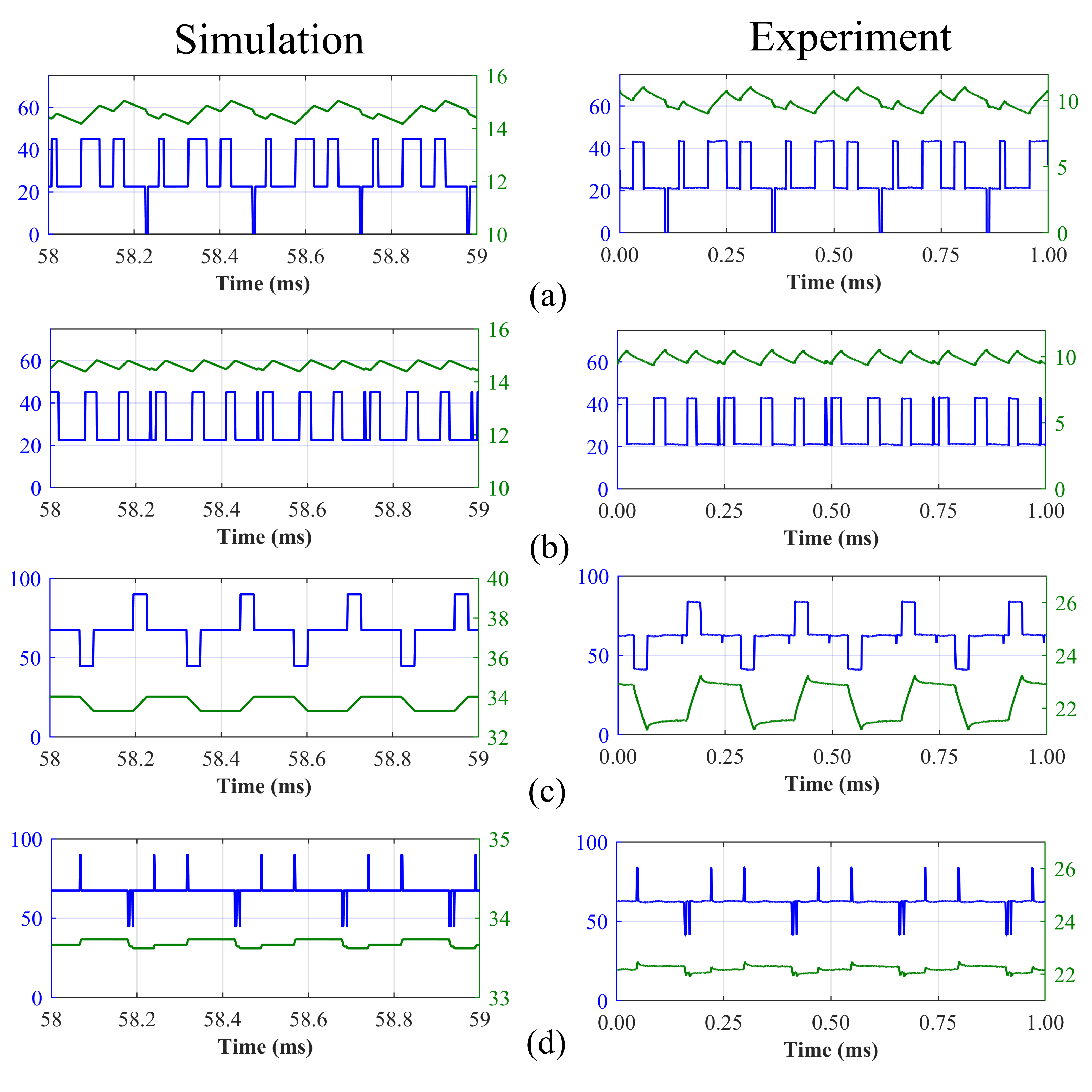}
    \caption{\footnotesize Battery pack output voltage and current.  
    Third scenario: (a) conventional PSC-PWM; (b) proposed method.  
    Fourth scenario: (c) conventional PSC-PWM; (d) proposed method.}
    \label{fig:11}
\end{figure}

Similarly, Figure~\ref{fig:10}(c) presents the pack voltage and current when the modulation indices are unevenly distributed above 0.5 (\(\mathbf{\textit{m}} = [0.55, 0.7, 0.8, 0.95]\)). As in the last case, the optimized phase shifts change the pulse width of the voltage steps to lower the WTHD and the current ripple (Figure~\ref{fig:10}(d)). Figure~\ref{fig:11}(a) illustrates the voltage and current of the pack under extreme unbalanced conditions (\(\mathbf{\textit{m}} = [0.2, 0.35, 0.6, 0.8]\)). With the proposed method, the current ripple and WTHD are notably reduced (Figure~\ref{fig:11}(b)). 

Finally, in a fault scenario where the modulation index of one module is set to zero (\(\mathbf{\textit{m}} = [0.4, 0.1, 0.25, 0]\)), which corresponds to real-world module failures (Figure~\ref{fig:11}(c)). The proposed method can perform well even under fault conditions (Figure~\ref{fig:11}(d)).

Table II provides a summary of the achieved results in the testbench. The proposed AI-driven approach outperforms conventional PSC-PWM in all four case studies, achieving up to 50\% reduction in current ripple in Case 1 and Case 3, and more than 50\% in Case 2 and Case 4. Moreover, WTHD reduction follows a similar trend.

\begin{table}[t]
\centering
\caption{Comparison of Current Ripple and WTHD for Different Modulation Conditions}
\label{tab:ripple_wthd_comparison}
\resizebox{0.48\textwidth}{!}{ % Adjust to fit within half-page width
\begin{tabular}{c | c c | c}
\hline \hline
\textbf{Case} & \textbf{Current Ripple (\%)} & \textbf{WTHD (\%)} & \textbf{Modulation Vector} (\(\mathbf{m}\)) \\ 
\hline
\multicolumn{4}{c}{\textbf{Scenario 1: All Modulation Indices Below 0.5}} \\ 
\hline
Conventional & \hfill 51.48 & \hfill 8.69 & \([0.15, 0.3, 0.4, 0.45]\) \\  
Proposed NN-Based & \hfill 26.78 & \hfill 5.97 & \([0.15, 0.3, 0.4, 0.45]\) \\  
\hline
\multicolumn{4}{c}{\textbf{Scenario 2: Unevenly Distributed Indices Above 0.5}} \\ 
\hline
Conventional & \hfill 12.29 & \hfill 1.76 & \([0.55, 0.7, 0.8, 0.95]\) \\  
Proposed NN-Based & \hfill 4.45 & \hfill 0.64 & \([0.55, 0.7, 0.8, 0.95]\) \\  
\hline
\multicolumn{4}{c}{\textbf{Scenario 3: Extreme Unbalanced Conditions}} \\ 
\hline
Conventional & \hfill 21.13 & \hfill 5.81 & \([0.2, 0.35, 0.6, 0.8]\) \\  
Proposed NN-Based & \hfill 12.67 & \hfill 2.73 & \([0.2, 0.35, 0.6, 0.8]\) \\  
\hline
\multicolumn{4}{c}{\textbf{Scenario 4: Fault Condition (One Module Fails)}} \\ 
\hline
Conventional & \hfill 9.39 & \hfill 1.70 & \([0.4, 0.1, 0.25, 0]\) \\  
Proposed NN-Based & \hfill 2.63 & \hfill 0.52 & \([0.4, 0.1, 0.25, 0]\) \\  
\hline
\end{tabular}
} % End of resizebox
\end{table}

\section{Conclusion}
We propose a novel scheduling framework suitable for modular multilevel converters, cascaded half-bridge converters, and any other assemblies whose fundamental structure is a string of modules. The proposed method combines PSC-PWM and neural networks to predict optimum phase-shift angles, which minimize current ripple and WTHD under imbalanced modulation indices and balanced modules. MATLAB\,/\,SIMULINK simulations and experimental tests on scaled-up reconfigurable batteries validate the performance and stability of the proposed method.

In contrast to existing techniques that rely on additional control loops or computationally demanding optimization algorithms, the proposed neural network requires only a single training phase for a specified number of modules and can dynamically adapt to fault or reduced-module operating conditions without retraining. Furthermore, a simple scaling strategy has been introduced, which allows a neural network trained for $N_b$ modules to be directly applied to scaled systems with $z \cdot N_b$ modules by dividing the modulation vector into sub-vectors and applying corresponding phase-shift offsets. This scalability eliminates the need for retraining. The avoidance of retraining establishes a high level of practicality of the proposed framework  for real-time modular converter applications.

\bibliographystyle{IEEEtran} % Ensures IEEE format
\bibliography{TII-Articles-LaTeX-template/reference} % Replace 'yourbibfile' with the actual name of your .bib file (without .bib extension)

@article{hosseini2023energy,
  title={Energy recovery and energy harvesting in electric and fuel cell vehicles, a review of recent advances},
  author={Hosseini, Seyed Mohammad and Soleymani, Mehdi and Kelouwani, Sousso and Amamou, Ali Akrem},
  journal={IEEE Access},
  volume={11},
  pages={83107--83135},
  year={2023},
  publisher={IEEE}
}

@article{shang2021fast,
  title={Fast grid frequency and voltage control of battery energy storage system based on the amplitude-phase-locked-loop},
  author={Shang, Lei and Dong, Xuzhu and Liu, Chengxi and Gong, Zhen},
  journal={IEEE Transactions on Smart Grid},
  volume={13},
  number={2},
  pages={941--953},
  year={2021},
  publisher={IEEE}
}

@article{chen2020unbalanced,
  title={Unbalanced operation principle and fast balancing charging strategy of a cascaded modular multilevel converter--bidirectional DC--DC converter in the shipboard applications},
  author={Chen, Peng and Xiao, Fei and Liu, Jilong and Zhu, Zhichao and Ren, Qiang},
  journal={IEEE Transactions on Transportation Electrification},
  volume={6},
  number={3},
  pages={1265--1278},
  year={2020},
  publisher={IEEE}
}

@ARTICLE{10436634,
  author={Hashemi-Zadeh, Amin and Ahmadi, Saeid and Neyshabouri, Yousef and Asadi, Ehsan and Iman-Eini, Hossein and Liserre, Marco},
  journal={IEEE Transactions on Power Electronics}, 
  title={An Enhanced Model Predictive Capacitor Voltage Control of Hybrid Modular Multilevel Converters Under Overmodulation Circumstances}, 
  year={2024},
  volume={39},
  number={6},
  pages={7130-7143},
  doi={10.1109/TPEL.2024.3365850}
}

@article{hossain2025advanced,
  title={Advanced PWM Technique for Performance Improvement of Solid-State Transformer-Based Renewable Energy Systems},
  author={Hossain, Md Sanwar and Islam, Md Rabiul and Biswas, Shuvra Prokash and Sutanto, Danny and Muttaqi, Kashem M},
  journal={IEEE Transactions on Industrial Electronics},
  year={2025},
  publisher={IEEE}
}

@article{du2025analysis,
  title={Analysis of Carrier Overlap Modulation and Space Vector Modulation Equivalence in NPC Inverters},
  author={Du, Haoting and He, Yingjie and Zhao, Zhengchen and Zhuang, Shenglun and Liu, Jinjun},
  journal={IEEE Transactions on Power Electronics},
  year={2025},
  publisher={IEEE}
}

@article{wang2021flexible,
  title={Flexible nearest level modulation for modular multilevel converter},
  author={Wang, Weiyao and Ma, Ke and Cai, Xu},
  journal={IEEE Transactions on Power Electronics},
  volume={36},
  number={12},
  pages={13686--13696},
  year={2021},
  publisher={IEEE}
}

@article{marquez2017variable,
  title={Variable-angle phase-shifted PWM for multilevel three-cell cascaded H-bridge converters},
  author={Marquez, Abraham and Leon, Jose I and Vazquez, Sergio and Portillo, Ramon and Franquelo, Leopoldo G and Freire, Emilio and Kouro, Samir},
  journal={IEEE Transactions on Industrial Electronics},
  volume={64},
  number={5},
  pages={3619--3628},
  year={2017},
  publisher={IEEE}
}

@ARTICLE{9275384,
  author={Tashakor, Nima and Kilictas, Muhammet and Bagheri, Ehsan and Goetz, Stefan},
  journal={IEEE Transactions on Power Electronics}, 
  title={Modular Multilevel Converter With Sensorless Diode-Clamped Balancing Through Level-Adjusted Phase-Shifted Modulation}, 
  year={2021},
  volume={36},
  number={7},
  pages={7725-7735},
  doi={10.1109/TPEL.2020.3041599}
}

@inproceedings{kacetl2022ageing,
  title={Ageing Mitigation and Loss Control in Reconfigurable Batteries in Series-Level Setups},
  author={Kacetl, Tomáš and Kacetl, Jan and Tashakor, Nima and Goetz, Stefan},
  booktitle={2022 24th European Conference on Power Electronics and Applications (EPE'22 ECCE Europe)},
  pages={1--10},
  year={2022},
  organization={IEEE},
  doi={10.23919/EPE22ECCEEurope.2022.9907223}
}

@article{tashakor2023generic,
  title={Generic Dynamically Reconfigurable Battery With Integrated Auxiliary Output and Balancing Capability},
  author={Tashakor, Nima and Kacetl, Jan and Fang, Jingyang and Li, Zhen and Goetz, Stefan},
  journal={IEEE Transactions on Power Electronics},
  volume={38},
  number={7},
  pages={7933--7944},
  year={2023},
  publisher={IEEE},
  doi={10.1109/TPEL.2023.3263809}
}

@inproceedings{goetz2016sensorless,
  title={Sensorless scheduling of the modular multilevel series-parallel converter: enabling a flexible, efficient, modular battery},
  author={Goetz, Stefan M. and Li, Zhen and Peterchev, Angel V. and Liang, Xue and Zhang, Cheng and Lukic, Srdjan M.},
  booktitle={2016 IEEE Applied Power Electronics Conference and Exposition (APEC)},
  pages={2349--2354},
  year={2016},
  organization={IEEE},
  doi={10.1109/APEC.2016.7468193}
}

@article{alcaide2021variable,
  title={Variable-Angle PS-PWM Technique for Multilevel Cascaded H-Bridge Converters With Large Number of Power Cells},
  author={Alcaide, A. M. and others},
  journal={IEEE Transactions on Industrial Electronics},
  volume={68},
  number={8},
  pages={6773--6783},
  year={2021},
  publisher={IEEE},
  doi={10.1109/TIE.2020.3000121}
}

@article{liu2021derivation,
  title={Derivation of the Generalized Phase-Shifted Angles by Using Phasor Diagrams for the CHB Converter With Unbalanced DC Voltage Sources},
  author={Liu, Peng and Duan, Shanxu},
  journal={IEEE Transactions on Industrial Electronics},
  volume={68},
  number={12},
  pages={12002--12009},
  year={2021},
  publisher={IEEE},
  doi={10.1109/TIE.2020.3040674}
}

@article{an2023selective,
  title={Selective Virtual Synthetic Vector Embedding for Full-Range Current Harmonic Suppression of the DC Collector},
  author={An, Feng and Zhao, Biao and Cui, Bin and Chen, Yushuo and Qu, Lu and Yu, Zhanqing and Zeng, Rong},
  journal={IEEE Transactions on Power Electronics},
  volume={38},
  number={2},
  pages={2577--2588},
  year={2023},
  publisher={IEEE},
  doi={10.1109/TPEL.2022.3209696}
}

@book{holmes2003pwm,
  author    = {Derek A. Holmes and Thomas A. Lipo},
  title     = {Pulse Width Modulation for Power Converters: Principles and Practice},
  year      = {2003},
  publisher = {Wiley-IEEE Press},
  isbn      = {978-0-471-20814-3},
  address   = {Piscataway, NJ},
  url       ={https://ieeexplore.ieee.org/document/5280117}
}

@article{zhao2020overview,
  title={An overview of artificial intelligence applications for power electronics},
  author={Zhao, Shuai and Blaabjerg, Frede and Wang, Huai},
  journal={IEEE Transactions on Power Electronics},
  volume={36},
  number={4},
  pages={4633--4658},
  year={2020},
  publisher={IEEE}
}

@article{jiao2021closed,
  title={The closed-loop sideband harmonic suppression for CHB inverter with unbalanced operation},
  author={Jiao, Ning and Wang, Shunliang and Ma, Junpeng and Chen, Xiang and Liu, Tianqi and Zhou, Dao and Yang, Yongheng},
  journal={IEEE Transactions on Power Electronics},
  volume={37},
  number={5},
  pages={5333--5341},
  year={2021},
  publisher={IEEE}
}

@ARTICLE{8576676,
  author={Zhang, Qinghao and Sun, Kai},
  journal={IEEE Journal of Emerging and Selected Topics in Power Electronics}, 
  title={A Flexible Power Control for PV-Battery Hybrid System Using Cascaded H-Bridge Converters}, 
  year={2019},
  volume={7},
  number={4},
  pages={2184-2195},
  doi={10.1109/JESTPE.2018.2887002}
}

@article{marquez2019generalized,
  title={Generalized harmonic control for CHB converters with unbalanced cells operation},
  author={Marquez, Abraham and Leon, Jose I and Monopoli, Vito Giuseppe and Vazquez, Sergio and Liserre, Marco and Franquelo, Leopoldo G},
  journal={IEEE Transactions on Industrial Electronics},
  volume={67},
  number={11},
  pages={9039--9047},
  year={2019},
  publisher={IEEE}
}

@article{hannan2021battery,
  title={Battery energy-storage system: A review of technologies, optimization objectives, constraints, approaches, and outstanding issues},
  author={Hannan, Mohammad Abdul and Wali, SB and Ker, Pin Jern and Abd Rahman, Muhamad Safwan and Mansor, M and Ramachandaramurthy, VK and Muttaqi, KM and Mahlia, Teuku Meurah Indra and Dong, Zhao Yang},
  journal={Journal of Energy Storage},
  volume={42},
  pages={103023},
  year={2021},
  publisher={Elsevier}
}

@article{boscaino2024grid,
  title={Grid-connected photovoltaic inverters: Grid codes, topologies and control techniques},
  author={Boscaino, Valeria and Ditta, Vito and Marsala, Giuseppe and Panzavecchia, Nicola and Tine, Giovanni and Cosentino, Valentina and Cataliotti, Antonio and Di Cara, Dario},
  journal={Renewable and Sustainable Energy Reviews},
  volume={189},
  pages={113903},
  year={2024},
  publisher={Elsevier}
}

@article{montesinos2012design,
  title={Design and control of a modular multilevel DC/DC converter for regenerative applications},
  author={Montesinos-Miracle, Daniel and Massot-Campos, Miquel and Bergas-Jane, Joan and Galceran-Arellano, Samuel and Rufer, Alfred},
  journal={IEEE transactions on power electronics},
  volume={28},
  number={8},
  pages={3970--3979},
  year={2012},
  publisher={IEEE}
}

@article{majeed2018multiple,
  title={A multiple-input cascaded DC--DC converter for very small wind turbines},
  author={Majeed, Yamaan E and Ahmad, Iftekhar and Habibi, Daryoush},
  journal={IEEE Transactions on Industrial Electronics},
  volume={66},
  number={6},
  pages={4414--4423},
  year={2018},
  publisher={IEEE}
}

@article{chen2022current,
  title={Current ripple mitigation strategy of modular multilevel dc/dc converter for battery energy storage system},
  author={Chen, Tianmu and Zeng, Guohong and Jing, Long and Wang, Sirui and Zhang, Weige},
  journal={IEEE Transactions on Industrial Electronics},
  volume={70},
  number={11},
  pages={11555--11565},
  year={2022},
  publisher={IEEE}
}

@article{monopoli2018improved,
  title={Improved harmonic performance of cascaded H-bridge converters with thermal control},
  author={Monopoli, Vito Giuseppe and Marquez, Abraham and Leon, Jose I and Ko, Youngjong and Buticchi, Giampaolo and Liserre, Marco},
  journal={IEEE Transactions on Industrial Electronics},
  volume={66},
  number={7},
  pages={4982--4991},
  year={2018},
  publisher={IEEE}
}

@inproceedings{meshram2024advancements,
  title={Advancements in Artificial Intelligence for Design, Control, and Maintenance of Power Converters: A Comprehensive Review},
  author={Meshram, Vipinkumar Shriram and Becchi, Lorenzo and Alfonso, Cristian Garzon and Paolucci, Libero and Grasso, Francesco and Reatti, Alberto},
  booktitle={2024 IEEE 8th Forum on Research and Technologies for Society and Industry Innovation (RTSI)},
  pages={506--511},
  year={2024},
  organization={IEEE}
}

@article{rajamony2022artificial,
  title={Artificial neural networks-based multi-objective design methodology for wide-bandgap power electronics converters},
  author={Rajamony, Rajesh and Wang, Sheng and Calderon-Lopez, Gerardo and Ludtke, Ingo and Ming, Wenlong},
  journal={IEEE Open Journal of Power Electronics},
  volume={3},
  pages={599--610},
  year={2022},
  publisher={IEEE}
}

@article{zhang2025topology,
  title={Topology, Control, and Applications of MMC with Embedded Energy Storage: A Brief Review},
  author={Zhang, Lidong and Zhu, Qionghai and Xiao, Huangqing},
  journal={Electronics},
  volume={14},
  number={5},
  pages={949},
  year={2025},
  publisher={MDPI}
}

@article{8000650,
  author={Li, Nan and Gao, Feng and Hao, Tianqu and Ma, Zhan and Zhang, Chenghui},
  journal={IEEE Transactions on Industrial Electronics}, 
  title={SOH Balancing Control Method for the MMC Battery Energy Storage System}, 
  year={2018},
  volume={65},
  number={8},
  pages={6581-6591},
  doi={10.1109/TIE.2017.2733462}
}

@INPROCEEDINGS{8341622,
  author={Ma, Zhan and Hao, Tianqu and Gao, Feng and Li, Nan and Gu, Xin},
  booktitle={2018 IEEE Applied Power Electronics Conference and Exposition (APEC)}, 
  title={Enhanced SOH balancing method of MMC battery energy storage system with cell equalization capability}, 
  year={2018},
  volume={},
  number={},
  pages={3591-3597},
  doi={10.1109/APEC.2018.8341622}
}

@article{leon2017multilevel,
  title={Multilevel converters: Control and modulation techniques for their operation and industrial applications},
  author={Leon, Jose I and Vazquez, Sergio and Franquelo, Leopoldo G},
  journal={Proceedings of the IEEE},
  volume={105},
  number={11},
  pages={2066--2081},
  year={2017},
  publisher={IEEE}
}

@article{Chuang2019PV,
  title={Concept of a distributed photovoltaic multilevel inverter with cascaded double H-bridge topology},
  author={Goetz, Stefan M. and Wang, Chuang and Li, Zhongxi and Murphy, David L. K. and Peterchev, Angel V.},
  journal={International Journal of Electrical Power \& Energy Systems},
  volume={110},
  pages={667--678},
  year={2019},
  publisher={Elsevier}
}

@article{kacetl2022bandwidth,
  title={Bandwidth-increased ripple-mitigating scheduling algorithm for dynamically reconfigurable batteries},
  author={Kacetl, Tom{\'a}{\v{s}} and Kacetl, Jan and Tashakor, Nima and Fang, Jingyang and Goetz, Stefan M},
  journal={IEEE Access},
  volume={10},
  pages={104202--104214},
  year={2022},
  publisher={IEEE}
}

@ARTICLE{8529274,
  author={Dekka, Apparao and Wu, Bin and Yaramasu, Venkata and Fuentes, Ricardo Lizana and Zargari, Navid R.},
  journal={IEEE Journal of Emerging and Selected Topics in Power Electronics}, 
  title={Model Predictive Control of High-Power Modular Multilevel Converters—An Overview}, 
  year={2019},
  volume={7},
  number={1},
  pages={168-183},
  doi={10.1109/JESTPE.2018.2880137}}

@article{wu2017high,
  author = {Wu, J. and Zhang, L.},
  title = {High-quality power conversion in advanced VSCs},
  journal = {IEEE Transactions on Power Electronics},
  volume = {32},
  number = {5},
  pages = {1234--1245},
  year = {2017}
}

@article{perez2021modular,
  author = {Perez, M. A. and Ceballos, S. and Konstantinou, G. and Pou, J. and Aguilera, R. P.},
  title = {Modular multilevel converters: Recent achievements and challenges},
  journal = {IEEE Open Journal of the Industrial Electronics Society},
  volume = {2},
  pages = {224--239},
  year = {2021}
}

@article{marquardt2018modular,
  author = {Marquardt, R.},
  title = {Modular multilevel converters: State of the art and future progress},
  journal = {IEEE Power Electronics Magazine},
  volume = {5},
  number = {4},
  pages = {24--31},
  year = {2018}
}

@article{rodriguez2002multilevel,
  author = {Rodriguez, J. and Lai, J.-S. and Peng, F. Z.},
  title = {Multilevel inverters: A survey of topologies, controls, and applications},
  journal = {IEEE Transactions on Industrial Electronics},
  volume = {49},
  number = {4},
  pages = {724--738},
  year = {2002},
  month = {Aug}
}

@article{li2020hybrid,
  title={Hybrid back-to-back MMC system for variable speed AC machine drives},
  author={Li, Binbin and Hu, Junlin and Zhou, Shaoze and Xu, Dianguo},
  journal={CPSS transactions on power electronics and applications},
  volume={5},
  number={2},
  pages={114--125},
  year={2020},
  publisher={CPSS}
}

@article{kumar2019balanced,
  title={Balanced submodule operation of modular multilevel converter-based induction motor drive for wide-speed range},
  author={Kumar, Yerraguntla Shasi and Poddar, Gautam},
  journal={IEEE Transactions on Power Electronics},
  volume={35},
  number={4},
  pages={3918--3927},
  year={2019},
  publisher={IEEE}
}

@inproceedings{hagiwara2012startup,
  title={Startup and low-speed operation of an adjustable-speed motor driven by a modular multilevel cascade inverter (MMCI)},
  author={Hagiwara, Makoto and Hasegawa, Isamu and Akagi, Hirofumi},
  booktitle={2012 IEEE energy conversion congress and exposition (ECCE)},
  pages={718--725},
  year={2012},
  organization={IEEE}
}

@inproceedings{okazaki2014research,
  title={Research trends of modular multilevel cascade inverter (MMCI-DSCC)-based medium-voltage motor drives in a low-speed range},
  author={Okazaki, Yuhei and Matsui, Hitoshi and Hagiwara, Makoto and Akagi, Hirofumi},
  booktitle={2014 International Power Electronics Conference (IPEC-Hiroshima 2014-ECCE ASIA)},
  pages={1586--1593},
  year={2014},
  organization={IEEE}
}

@article{flourentzou2009overview,
  author = {Flourentzou, N. and Agelidis, V. G. and Demetriades, G. D.},
  title = {VSC-based HVDC power transmission systems: An overview},
  journal = {IEEE Transactions on Power Electronics},
  volume = {24},
  number = {3},
  pages = {592--602},
  year = {2009},
  month = {Mar}
}

@article{ansari2020mmc,
  title={MMC based MTDC grids: A detailed review on issues and challenges for operation, control and protection schemes},
  author={Ansari, Jamshed Ahmed and Liu, Chongru and Khan, Shahid Aziz},
  journal={IEEE Access},
  volume={8},
  pages={168154--168165},
  year={2020},
  publisher={IEEE}
}

@article{sun2022beyond,
  title={Beyond the MMC: Extended modular multilevel converter topologies and applications},
  author={Sun, Pingyang and Tian, Yumeng and Pou, Josep and Konstantinou, Georgios},
  journal={IEEE Open Journal of Power Electronics},
  volume={3},
  pages={317--333},
  year={2022},
  publisher={IEEE}
}

@article{li2018operation,
  title={Operation and control methods of modular multilevel converters in unbalanced AC grids: A review},
  author={Li, Jinke and Konstantinou, Georgios and Wickramasinghe, Harith R and Pou, Josep},
  journal={IEEE Journal of Emerging and Selected Topics in Power Electronics},
  volume={7},
  number={2},
  pages={1258--1271},
  year={2018},
  publisher={IEEE}
}

@book{sharifabadi2016design,
  title={Design, control, and application of modular multilevel converters for HVDC transmission systems},
  author={Sharifabadi, Kamran and Harnefors, Lennart and Nee, Hans-Peter and Norrga, Staffan and Teodorescu, Remus},
  year={2016},
  publisher={John Wiley \& Sons}
}

@article{goetz2015modular,
  author = {Goetz, S. M. and Peterchev, A. V. and Weyh, T.},
  title = {Modular multilevel converter with series and parallel submodule connectivity: Topology and control},
  journal = {IEEE Transactions on Power Electronics},
  volume = {30},
  number = {1},
  pages = {203--215},
  year = {2015},
  month = {Jan}
}

@article{cheung2018transformerless,
  author    = {V. S. Cheung and R. S. Yeung and H. S. Chung and A. W. Lo and W. Wu},
  title     = {A transformer-less unified power quality conditioner with fast dynamic control},
  journal   = {IEEE Transactions on Power Electronics},
  volume    = {33},
  number    = {5},
  pages    = {3926--3937},
  year      = {2018},
  month     = {May},
  doi       = {10.1109/TPEL.2017.2737026}
}

@article{Wang2019PV,
  title={Concept of a distributed photovoltaic multilevel inverter with cascaded double H-bridge topology},
  author={Goetz, Stefan M and Wang, Chuang and Li, Zhongxi and Murphy, David LK and Peterchev, Angel V},
  journal={International Journal of Electrical Power \& Energy Systems},
  volume={110},
  pages={667--678},
  year={2019}
}

@misc{Divide,
    author = {Goetz, Stefan M},
    title = {Method for operating a modular multilevel converter},
    year = {2016},
    note = {{DE}\,10\,2016\,109\,077, Deutsches Patent- und Markenamt}
}

@article{wang2025selective,
  title={Selective Harmonic Elimination-Based Torque Ripple Optimization for PMSMs with Model Predictive Pulse Pattern Control},
  author={Wang, Chenxu and Zhang, Qi and Yang, Hansen and Wu, Mingzhe and Wang, Kui and Li, Yongdong and Ge, Shirong and Yang, Kehu},
  journal={IEEE Transactions on Power Electronics},
  year={2025},
  publisher={IEEE}
}

@article{chen2025analysis,
  title={Analysis of interactions in offshore wind farm transmission system utilizing DR-MMC rectifier},
  author={Chen, Yin and Chen, Yuntao and Yang, Ning and Xu, Lie and Egea-{\`A}lvarez, Agust{\'\i} and Chen, Xia},
  journal={IEEE Transactions on Power Delivery},
  year={2025},
  publisher={IEEE}
}
\end{document}